\documentclass[aps,preprint,nofootinbib,superscriptaddress,prc]{revtex4}
\usepackage{amssymb}

\usepackage{epsfig}
\usepackage[bookmarksnumbered,bookmarksopen,colorlinks,citecolor=blue,linkcolor=blue]{hyperref}
\begin{document}


\title{ Correlations between neutrons and protons near Fermi surface and $Q_{\alpha}$ of super-heavy nuclei}
\author{Ning Wang }
\thanks{wangning@gxnu.edu.cn}
\affiliation{Department of Physics, Guangxi Normal University,
Guilin 541004, P. R. China}
\affiliation{ State Key Laboratory of
Theoretical Physics, Institute of Theoretical Physics, Chinese
Academy of Sciences, Beijing 100190, People's Republic of China}

\author{Min Liu}
\affiliation{Department of Physics, Guangxi Normal University,
Guilin 541004, P. R. China}

\author{Xizhen Wu}
\affiliation{China Institute of Atomic Energy, Beijing 102413, P.
R. China}

\author{Jie Meng}
\affiliation{State Key Laboratory of Nuclear Physics and Technology,
School of Physics, Peking University, Beijing 100871, China}

\begin{abstract}
The shell corrections and shell gaps in nuclei are systematically
studied with the latest Weizs\"acker-Skyrme (WS4) mass model. We find
that most of asymmetric nuclei with (sub)-shell closures locate
along the shell stability line (SSL), $N=1.37Z+13.5$, which might be
due to a strong correlation between neutrons and
protons near Fermi surface. The double magicity of nuclei $^{46}$Si and $^{78}$Ni is
predicted according to the corresponding shell gaps, shell
corrections and nuclear deformations. The unmeasured super-heavy nuclei $^{296}$118 and
$^{298}$120, with relatively large shell gaps and shell corrections, also locate
along the SSL, whereas the traditional magic nucleus $^{298}$Fl
evidently deviates from the line. The $\alpha$-decay energies of
super-heavy nuclei with $Z=113-126$ are simultaneously investigated
by using the WS4 model together with the radial basis function
corrections. For super-heavy nuclei with large shell corrections, the smallest $\alpha$-decay energy for elements $Z=116$, 117  and 118 in their isotope chains locates at $N=178$ rather than $184$.

\end{abstract}

\maketitle

\begin{center}
\textbf{I. INTRODUCTION}
\end{center}

For nuclear physics, one of the most important tasks is to explore
the nuclear landscape. Up to now, the masses of 2438 nuclei have
already been measured according to the latest nuclear mass database
AME2012 \cite{Audi12}, and about 4000 $\sim$ 5000 masses are still
unknown. The masses of these unmeasured nuclei play a key role for the study of super-heavy nuclei (SHN) \cite{Sob,Ogan,Xu13}, the
r-process in nuclear astrophysics \cite{Meng,Mend14,Mum15} and nuclear symmetry energy \cite{Liu10,Wang15a,Zhao10}. For
the synthesis of SHN, the first question that should
be answered is where the central position of the island of stability
locates. Traditionally, the island of stability for SHN is predicted
to be around neutron number $N=184$ and proton number $Z=114$ \cite{Sob66}, 120 or 126 \cite{Hof00,Zeld98,Long14},
according to the large shell corrections (in absolute value) and/or the large
shell gaps in super-heavy nuclei, since the survival of these nuclei is directly due to the shell effects. From the predicted evaporation residue cross-sections, Adamian et al. concluded that $Z=114$ is not a proper magic number and the next magic nucleus beyond $^{208}$Pb is the nucleus with $Z\ge 120$ \cite{Adam09}. The uncertainty of model parameters and the decreasing trend of the shell gaps with increasing of nuclear size cause some difficulties in the determination of the central position of the island. The improvement of model predictive power and the investigation of the physics behind model parameters are helpful for the determination of the island and of the new magic numbers in extremely neutron-rich nuclei.

In addition to the properties of nuclear force represented by the model parameters, the concept of symmetry in physics is a very powerful tool for
understanding the behavior of Nature. The isospin symmetry discovered by Heisenberg plays an important role in nuclear
physics. In the absence of Coulomb interactions between the
protons, a perfectly charge-symmetric and charge-independent
nuclear force would result in the binding energies of mirror
nuclei being identical \cite{Lenzi,Shlomo}. For neutron-rich nuclei around the doubly-magic nuclei, the correlation between valence-nucleons and the corresponding doubly-magic core, and as well as the correlation/symmetry between active-protons and active-neutrons near Fermi surface should also affect the nuclear masses. It is therefore necessary to investigate these correlations.

For the synthesis of SHN, the $\alpha$-decay chain is a key quantity to identify the produced SHN. The masses of SHN directly influence the evaluation of the corresponding $\alpha$-decay energy $Q_\alpha$. Inspired by the Skyrme energy-density functional, a macroscopic-microscopic mass model, Weizs\"acker-Skyrme (WS) model \cite{Wang,Wang10,Liu11,Wang14} was proposed. In this model, the isospin dependence of model parameters and the mirror corrections from the isospin symmetry in nuclear physics play a crucial role for improving the accuracy of mass predictions for neutron-rich nuclei and super-heavy nuclei. By adopting the latest version (WS4) of the model \cite{Wang14} together with the radial basis function corrections \cite{Wang11, Niu13} which is a prominent global interpolation and extrapolation scheme to effectively describe the systematic error of a global mass model, the 2353 measured masses ($N\ge 8$ and $Z\ge 8$) in 2012 Atomic Mass Evaluation (AME) can be reproduced with an rms deviation of 170 keV. With an accuracy smaller than 300 keV for the $Q_\alpha$ of SHN, the WS4$^{\rm RBF}$ model is one of the most reliable mass models for the study of SHN \cite{Sob13,Lit14,Ogan15}. It is therefore interesting to systematically investigate the surface of $Q_\alpha$ in super-heavy region with this model.

In this work, we first study the shell corrections and shell gaps in unmeasured neutron-rich nuclei and super-heavy nuclei by using the WS4$^{\rm RBF}$ model. Simultaneously, the relationship among known doubly-magic nuclei such as $^{132}$Sn, $^{208}$Pb and $^{270}$Hs \cite{Sob,Ogan} will be analyzed from the point of view of the correlations between nucleons near Fermi surface. Then, the $Q_\alpha$ of super-heavy nuclei around the possible central position of the island of stability will be predicted.

\begin{center}
\textbf{II. SHELL CORRECTIONS AND SHELL GAPS IN NUCLEI}
\end{center}

In the Weizs\"acker-Skyrme mass model, the shell correction is obtained with the traditional Strutinsky's procedure in which the single-particle levels are calculated from an axially deformed Woods-Saxon potential \cite{Cwoik}.  In Fig. 1, we show the contour plot of the calculated shell corrections from the WS4 model for nuclei over the whole nuclear landscape. Obviously, the shell corrections (in absolute value) for the known doubly-magic nuclei such as $^{132}$Sn, $^{208}$Pb and $^{270}$Hs are larger than those of their neighboring nuclei. In addition to these known doubly-magic nuclei, the shell corrections for $^{46}$Si and $^{78}$Ni are also evidently large. $^{78}$Ni could be doubly-magic nucleus since both the proton number $Z=28$ and neutron number $N=50$ are magic numbers in known mass region. Very recently, the experimental investigation on the single-neutron states in $^{79}$Zn at CERN supports the picture of a robust $N=50$ shell closure for $^{78}$Ni \cite{Orlandi}.  For $^{46}$Si, the neutron number $N=32$ could be new magic number, since both the measured large shell gap in $^{52}$Ca \cite{Wien13} and the calculations from the WS* model \cite{Wang15} indicate $N=32$ being a magic number in neutron-rich nuclei. Some investigations suggested that $Z=14$ could also be proton magic number \cite{Myer66,Jan05,Ang15}. The neutron separation energy of Silicon and Nickel isotopes are also calculated with the WS4$^{\rm RBF}$ model, and the results are presented in Fig. 2. The squares and circles denote the measured neutron ($S_n$) and two-neutron ($S_{2n}$) separation energies, respectively, which can be remarkably well reproduced by the model predictions (the curves). According to the predicted neutron separation energy, $^{46}$Si could be the neutron drip-line nucleus and $^{78}$Ni is a well bound nucleus comparing with $^{46}$Si. The shell gaps in $^{46}$Si and $^{78}$Ni will be discussed later.

\begin{figure}
\includegraphics[angle=-0,width= 0.75\textwidth]{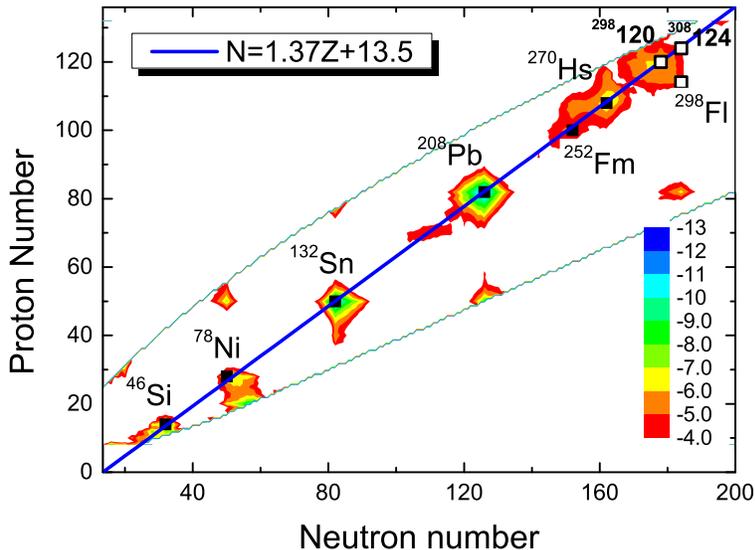}
 \caption{(Color online) Shell corrections of nuclei from the WS4 calculations.
 The solid and open squares denote the positions of nuclei with (sub)-shell closure according to the predicted shell gaps.      }
\end{figure}

\begin{figure}
\includegraphics[angle=-0,width= 0.75\textwidth]{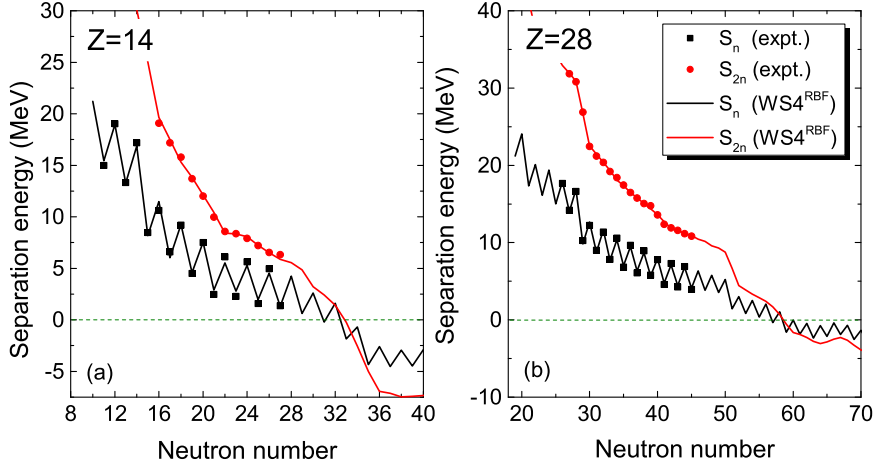}
 \caption{(Color online) Neutron separation energy of Silicon and Nickel isotopes. The squares and circles denote the measured neutron ($S_n$) and two-neutron ($S_{2n}$) separation energies, respectively. The curves denote the predictions of the WS4$^{\rm RBF}$ model. }
\end{figure}

More interestingly, one can see from Fig. 1 that the asymmetric nuclei with large shell corrections locate along the straight line $N=1.37Z+13.5$, with an uncertainty of neutron number $\Delta N<2$. To explore the physics behind this line, we study the correlation between neutrons and protons near the Fermi surface in these doubly-magic nuclei. It is known that nucleons near the Fermi surface can significantly influence the properties of nuclei, whereas the influence from an individual nucleon located at the deep part of potential well might be negligible. In this work, the nucleons in the same major shell which is nearest to the Fermi surface is defined as active nucleons, and the rest of nucleons form a relatively inactive core. As an example, the structure of $^{208}$Pb could be described as the core with $N=82$ and $Z=50$ together with the active-nucleons near the Fermi surface. The ratio of active-neutrons to active-protons is $N_a/Z_a=(126-82)/(82-50)=1.375$, and the isospin asymmetry of the core $I_{\rm core}=(82-50)/132=0.242$. Here, we introduce an effective ratio $T_{\rm eff}=N_a/Z_a-I_{\rm core}$. We find that one gets almost the same value $T_{\rm eff}=1.17\pm 0.06$ for all these doubly-magic nuclei. For symmetric nuclei, one gets $T_{\rm eff}=1$ due to the isospin symmetry. For heavy nuclei with $N>Z$, more neutrons are required to balance the strong Coulomb repulsion and the effective ratio $T_{\rm eff}$ should be larger than one. The similar value of $T_{\rm eff}$ indicates that there exists a strong correlation between the neutrons and protons near the Fermi surface. This correlation could be due to the competition between the isospin symmetry and Coulomb interaction in the active-nucleon space. The line $N=1.37Z+13.5$ which is called shell stability line (SSL) and the similar $T_{\rm eff}$ value imply that the symmetry in active-nucleon space could also influence the binding energies of asymmetric nuclei.

\begin{figure}
\includegraphics[angle=-0,width=0.7\textwidth]{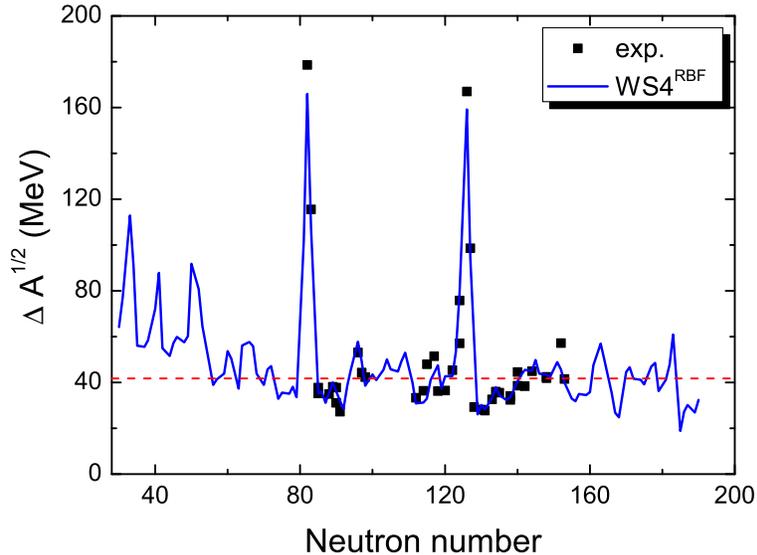}
 \caption{(Color online) Scaled shell gaps of nuclei along the shell stability line. The squares and solid curve denote
 the experimental data and WS4$^{\rm RBF}$ predictions, respectively. The dashed line denotes the mean value of measured shell gaps for all known nuclei. }
\end{figure}

\begin{table}
\caption{ Nuclei with relatively large shell gaps around the shell
stability line. $\Delta_{\rm scale}$ denotes the scaled shell gap
(in MeV) from the WS4$^{\rm RBF}$ predictions. $\Delta E$ and
$\beta_2$ denotes the corresponding shell correction (in MeV) and
the quadrupole deformation of nucleus according to the WS4 model,
respectively. }
\begin{tabular}{cccr}
 \hline\hline
   Nuclide   & ~~~$\Delta_{\rm scale}$~~~  & ~~~$\Delta E$~~~ & ~~~$\beta_2$ \\
\hline
 $^{46}$Si       & $105.1$ & $-7.96$ &  $-0.01 $\\
 $^{60}$Ca       &  $97.9$ & $-1.32$ &  $-0.01 $\\
 $^{78}$Ni       & $115.1$ & $-7.98$ &  $ 0.01 $\\
 $^{132}$Sn       &$166.0$ & $-12.10$&  $ 0.01 $\\
 $^{208}$Pb       &$159.2$ & $-12.43$&  $ 0.00 $\\
 $^{252}$Fm       &$55.3$  & $-5.30$ &  $ 0.24 $\\
 $^{270}$Hs       &$61.1$  & $-6.95$ &  $ 0.22 $\\
 $^{296}$118      &$48.0$  & $-5.93$ &  $-0.08 $\\
 $^{298}$120      &$48.6$  & $-5.89$ &  $-0.08 $\\
 $^{308}$124      &$67.0$  & $-4.31$ &  $ 0.00 $\\

 \hline\hline
\end{tabular}
\end{table}

To understand the magic numbers in extremely neutron-rich region and super-heavy region, we simultaneously investigate the shell gaps in nuclei.
In macroscopic-microscopic models, the shell correction provides a natural measure for magicity. A more direct measure of a shell closure is the
observation of a sudden jump in the two-nucleon separation energies \cite{Rutz}. The empirical shell gaps in nuclei are defined as the sum of the neutron and proton shell gaps based on the difference of the binding energies (in negative values) of nuclei,
\begin{eqnarray}
\Delta (N,Z)=\Delta_n (N,Z)+\Delta_p (N,Z),
\end{eqnarray}
with
\begin{eqnarray}
\Delta_n (N,Z)=B(N+2,Z)+B(N-2,Z)-2B(N,Z)
\end{eqnarray}
and
\begin{eqnarray}
\Delta_p (N,Z)=B(N,Z+2)+B(N,Z-2)-2B(N,Z).
\end{eqnarray}
Usually, the two-nucleon gaps show a pronounced peak for magic numbers and can be considered as indicators of a shell closure \cite{Rutz}. Here, we would like to emphasize that the large shell gaps in light nuclei with $N=Z$ is partly due to the Wigner effects which was evidently observed in \cite{Wang14b}. In this work, we focus on the shell gaps in nuclei with $N>Z$. At the same time, we introduce a scaled shell gap $\Delta_{\rm scale} (N,Z)=\Delta (N,Z) A^{1/2}$ in order to study the change of $\Delta (N,Z)$ with the simlar scale for both light and heavy nuclei \cite{Wang14b}. The mean value of measured shell gaps for known nuclei is $\langle \Delta_{\rm scale}\rangle=41.9$ MeV. The corresponding value for 39 known nuclei around the SSL is $\langle \Delta^{\rm SSL}_{\rm scale}\rangle=50.4$ MeV, which is obviously larger than the mean value for all known nuclei. The predicted mean value for the 118 nuclei along the SSL is $\langle \Delta^{\rm SSL}_{\rm scale}\rangle ({\rm WS4}^{\rm RBF})=49.1$ MeV. The calculated scaled shell gaps in nuclei around the SSL are also shown in Fig. 3 as a function of neutron number. The measured shell gaps can be remarkably well reproduced by the WS4$^{\rm RBF}$ calculations. The dashed line denotes the mean value for all known nuclei. The peaks that evidently larger than the mean value $\langle \Delta_{\rm scale}\rangle$ might imply the appearance of (sub)-shell closures in the corresponding nuclei. In Table I, we list some nuclei with relatively large shell gaps around the SSL from the WS4$^{\rm RBF}$ predictions. The corresponding shell corrections $\Delta E$ and quadrupole deformations $\beta_2$ from the WS4 calculations are also presented. From the table one sees that both the predicted shell gaps and shell corrections (in absolute value) are very large for $^{46}$Si and $^{78}$Ni. Simultaneously, these two nuclei are generally spherical in shape according to the predicted ground state deformations. These calculations indicate that $^{46}$Si and $^{78}$Ni are doubly-magic nuclei. For SHN around the SSL, both the shell corrections (in absolute value) and shell gaps in $^{296}$118 and $^{298}$120 are relatively large.

\begin{center}
\textbf{III. $\alpha$-DECAY ENERGIES OF SUPER-HEAVY NUCLEI}
\end{center}

\begin{figure}
\includegraphics[angle=-0,width= 0.8\textwidth]{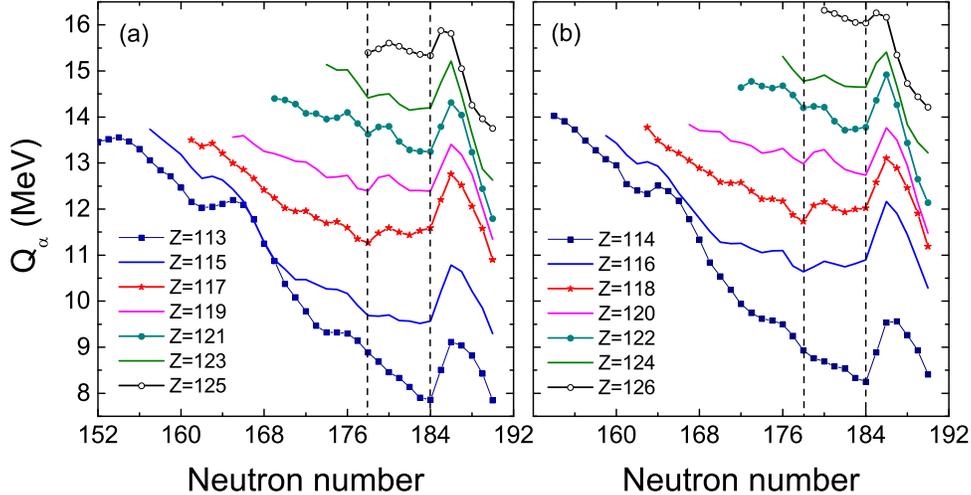}
 \caption{(Color online) $\alpha$-decay energies of odd-$Z$ super-heavy nuclei (a) and those of even-$Z$ nuclei (b) from the WS4$^{\rm RBF}$ predictions.       }
\end{figure}

In this work, the $\alpha$-decay energies $Q_\alpha$ SHN are systematically investigated with the WS4$^{\rm RBF}$ model. Previously, Oganessian and Utyonkov investigated the discrepancy between theory and experiment in $\alpha$-decay energies $\Delta Q_\alpha=Q^{\rm exp}_\alpha - Q^{\rm th}_\alpha$ for all of the nuclei produced as evaporation residues in the $^{48}$Ca-induced reactions and their daughter products. It is found that for all of the nuclei, including odd-$N$ and/or odd-$Z$ ones, the discrepancies $\Delta Q_\alpha$  are within $+0.5$ to $-0.4$ MeV from the WS4$^{\rm RBF}$ calculations \cite{Ogan}.
In Table II and Table III, we list the $\alpha$-decay chains for some SHN with even-$Z$. The corresponding shell corrections and deformation energies $E_{\rm def}$ (the difference in energy of a nucleus between its spherical and equilibrium shapes \cite{Sob}) are also presented. The predicted $\alpha$-decay energies based on the ground state energies of nuclei from four macroscopic-microscopic models are also listed for comparison. From the tables, one sees that for the considered SHN with $Z\le 118$, the predicted $Q_\alpha$ from the four different macroscopic-microscopic models are close to each other. For element 120, the results of the microscopic-macroscopic calculations based on the two-center shell model (TCSM) are evidently smaller than those of the other three models.

The predicted $Q_\alpha$ of SHN from the WS4$^{\rm RBF}$ calculations are simultaneously shown in Fig. 4. The dashed lines denote the positons of $N=178$ and 184. For nuclei with $Z\le 115$ and $N<186$, the smallest $Q_\alpha$ locates at $N=184$.  Whereas, for SHN with $116 \le Z \le 118$ and $N<186$, the smallest $Q_\alpha$ also locates at $N=178$.  For SHN with $Z=120$, there are two minima for the $\alpha$-decay energy: $Q_\alpha=12.98$ MeV at $N=178$ which is along the SSL (see Fig. 1) and $Q_\alpha=12.74$ MeV at $N=184$. For the already synthesized SHN $^{294}$118 through hot fusion reaction $^{48}$Ca+$^{249}$Cf \cite{Ogan06}, the neutron number $N=176$ is very close to the position of $N=178$. It is therefore very interesting and important to produce more neutron-rich SHN such as $^{296}$118 and $^{297}$118 to check the trend of $Q_\alpha$ with neutron number, since the predicted smallest $\alpha$-decay energy $Q_\alpha=11.73$ MeV for element $Z=118$ locates at $N=178$. The predicted quadrupole deformation $\beta_2=-0.08$ (see Table I) indicates that $^{296}$118 is not exactly spherical at its ground state. The deformation energy of 0.76 MeV indicates that $^{296}$118 is more stable with a slightly oblate shape in the fission path, since an extra-energy is required from oblate shape to the saddle point (with prolate shapes) and the fission path is longer comparing with the case from spherical shape. In addition, one can see from Fig. 4 that the smallest $Q_\alpha$ for element $Z \ge 120$ locates at $N=184$ again. Interestingly, we find that the nucleus $^{308}$124 locates along the SSL and the corresponding shell gap is also large. These investigations indicate that the corrections between active-neutrons and active-protons influence not only the shell gaps but also the $\alpha$-decay energies of SHN.

\begin{table}
\caption{ Shell corrections $\Delta E$, deformation
energies $E_{\rm def}$ from the WS4 calculations and $\alpha$-decay energies $Q_{\alpha}$ of some even-even SHN (in MeV). $Q_{\alpha}^{\rm WS}$ denotes the predicted $Q_{\alpha}$ from the WS4$^{\rm RBF}$ model. $Q_{\alpha}^{\rm MM}$ denotes the $Q_{\alpha}$ of the macroscopic-microscopic calculations in Refs. \cite{Mun03,Mun03a}. $Q_{\alpha}^{\rm TCSM}$ and $Q_{\alpha}^{\rm FRDM}$ denotes the results of two-center shell model (TCSM) \cite{Kuz12} and of the finite range droplet (FRDM) model \cite{Moll95}, respectively. The experimental $Q_{\alpha}^{\rm exp}$ are also presented. }
\begin{tabular}{cccccccccc}
 \hline\hline
 ~~~$A$~~~&~~~$Z$~~~  & ~~~$\Delta E$~~~ & ~~~$E_{\rm def}$~~~ & ~~~$Q_{\alpha}^{\rm WS}$~~~& ~~~$Q_{\alpha}^{\rm MM}$~~~& ~~~$Q_{\alpha}^{\rm TCSM}$~~~& ~~~$Q_{\alpha}^{\rm FRDM}$~~~& ~~~$Q_{\alpha}^{\rm exp}$~~~\\
\hline
 $308$  & $126$   & $-3.41$ &  $ 0.07$ & $16.14$ & $-$ & $13.33$ & $-$ & $-$ \\
 $304$  & $124$   & $-4.79$ &  $ 0.07$ & $14.91$ & $-$ & $13.50$ & $13.64$ & $-$ \\
 $300$  & $122$   & $-5.36$ &  $ 0.30$ & $14.20$ & $-$ & $13.62$ & $13.99$ & $-$ \\
 $296$  & $120$   & $-6.23$ &  $ 1.02$ & $13.32$ & $13.23$ & $11.78$ & $13.69$ & $-$ \\
 $292$  & $118$   & $-6.02$ &  $ 1.74$ & $12.21$ & $12.15$ & $12.03$ & $12.37$ & $-$ \\
 $288$  & $116$   & $-5.27$ &  $ 1.36$ & $11.26$ & $11.54$ & $10.92$ & $11.32$ & $-$ \\
 $284$  & $114$   & $-4.51$ &  $ 0.47$ & $10.54$ & $11.53$ & $10.71$ & $ 9.44$ & $-$ \\
 \\						
 $310$  & $126$   & $-3.23$ &  $ 0.00$ & $16.04$ & $-$ & $13.09$ & $-$ & $-$ \\
 $306$  & $124$   & $-4.64$ &  $ 0.15$ & $14.67$ & $-$ & $12.84$ & $16.33$ & $-$ \\
 $302$  & $122$   & $-5.08$ &  $ 0.25$ & $14.21$ & $-$ & $12.76$ & $14.05 $ & $-$ \\
 $298$  & $120$   & $-5.89$ &  $ 0.45$ & $12.98$ & $13.44$ & $11.33$ & $13.36$ & $-$ \\
 $294$  & $118$   & $-5.75$ &  $ 1.24$ & $12.17$ & $12.11$ & $11.53$ & $12.28$ & $11.81(6)$ \cite{PRC70}\\
 $290$  & $116$   & $-5.53$ &  $ 1.39$ & $11.06$ & $11.08$ & $10.90$ & $11.12$ & $11.00(8)$ \cite{PRC70}\\
 $286$  & $114$   & $-4.72$ &  $ 0.67$ & $ 9.94$ & $10.86$ & $10.38$ & $ 9.40 $ & $10.35(6)$ \cite{PRC70}\\
\\						
 $312$  & $126$   & $-2.66$ &  $ 1.24$ & $16.16$ & $-$ & $14.36$ & $16.53$ & $-$ \\
 $308$  & $124$   & $-4.31$ &  $ 0.00$ & $14.64$ & $-$ & $12.37$ & $16.14$ & $-$ \\
 $304$  & $122$   & $-5.05$ &  $ 0.33$ & $13.71$ & $-$ & $11.98$ & $14.82$ & $-$ \\
 $300$  & $120$   & $-5.27$ &  $ 0.18$ & $13.29$ & $13.11$ & $11.09$ & $13.40$ & $-$ \\
 $296$  & $118$   & $-5.93$ &  $ 0.76$ & $11.73$ & $12.06$ & $11.01$ & $12.29$ & $-$ \\
 $292$  & $116$   & $-5.33$ &  $ 0.87$ & $11.10$ & $11.06$ & $10.77$ & $10.83$ & $10.80(7)$ \cite{PRC70}\\
 $288$  & $114$   & $-5.04$ &  $ 0.78$ & $ 9.62$ & $10.32$ & $10.33$ & $ 9.17 $ & $10.09(7)$ \cite{PRC70}\\

 \hline\hline
\end{tabular}
\end{table}

\begin{table}
\caption{ The same as Table II, but for Odd-A nuclei. }
\begin{tabular}{cccccccccc}
 \hline\hline
 ~~~$A$~~~&~~~$Z$~~~  & ~~~$\Delta E$~~~ & ~~~$E_{\rm def}$~~~ & ~~~$Q_{\alpha}^{\rm WS}$~~~& ~~~$Q_{\alpha}^{\rm MM}$~~~& ~~~$Q_{\alpha}^{\rm TCSM}$~~~& ~~~$Q_{\alpha}^{\rm FRDM}$~~~& ~~~$Q_{\alpha}^{\rm exp}$~~~\\
\hline
 $309$  & $126$   & $-3.26$ &  $ 0.02$ & $16.05$ & $-$ & $13.21$ & $-$ & $-$ \\
 $305$  & $124$   & $-4.70$ &  $ 0.14$ & $14.77$ & $-$ & $13.00$ & $13.44$ & $-$ \\
 $301$  & $122$   & $-5.14$ &  $ 0.23$ & $14.23$ & $-$ & $13.21$ & $13.90$ & $-$ \\
 $297$  & $120$   & $-5.79$ &  $ 0.71$ & $13.12$ & $13.49$ & $11.53$ & $13.54$ & $-$ \\
 $293$  & $118$   & $-6.01$ &  $ 1.52$ & $12.21$ & $11.93$ & $11.69$ & $12.28$ & $-$ \\
 $289$  & $116$   & $-5.35$ &  $ 1.41$ & $11.15$ & $11.22$ & $10.85$ & $11.27$ & $-$ \\
 $285$  & $114$   & $-4.57$ &  $ 0.58$ & $10.25$ & $11.11$ & $10.52$ & $ 9.35$ & $10.56(5)$ \cite{PRC92}\\
\\						
 $311$  & $126$   & $-2.87$ &  $ 0.42$ & $16.26$ & $-$ & $13.84$ & $17.08$ & $-$ \\
 $307$  & $124$   & $-4.33$ &  $ 0.03$ & $14.66$ & $-$ & $12.54$ & $16.06$ & $-$ \\
 $303$  & $122$   & $-5.05$ &  $ 0.33$ & $13.91$ & $-$ & $12.22$ & $14.71$ & $-$ \\
 $299$  & $120$   & $-5.48$ &  $ 0.24$ & $13.23$ & $13.23$ & $11.23$ & $13.11$ & $-$ \\
 $295$  & $118$   & $-5.85$ &  $ 1.01$ & $11.88$ & $12.22$ & $11.25$ & $12.19$ & $-$ \\
 $291$  & $116$   & $-5.37$ &  $ 1.16$ & $11.09$ & $10.91$ & $10.75$ & $11.12$ & $10.89(7)$ \cite{PRC70}\\
 $287$  & $114$   & $-4.83$ &  $ 0.76$ & $ 9.74$ & $10.56$ & $10.31$ & $ 9.31$ & $10.16(6)$ \cite{PRC70}\\
\\						
 $313$  & $126$   & $-2.42$ &  $ 2.03$ & $15.34$ & $-$ & $14.45$ & $15.97$ & $-$ \\
 $309$  & $124$   & $-3.18$ &  $ 0.06$ & $15.17$ & $-$ & $13.16$ & $16.49$ & $-$ \\
 $305$  & $122$   & $-4.59$ &  $ 0.12$ & $13.74$ & $-$ & $11.35$ & $14.94$ & $-$ \\
 $301$  & $120$   & $-5.14$ &  $ 0.21$ & $13.04$ & $13.11$ & $10.98$ & $13.67$ & $-$ \\
 $297$  & $118$   & $-5.58$ &  $ 0.57$ & $12.08$ & $11.91$ & $10.88$ & $12.11$ & $-$ \\
 $293$  & $116$   & $-5.61$ &  $ 0.77$ & $10.77$ & $10.09$ & $10.51$ & $10.94$ & $10.67(6)$ \cite{PRC70}\\
 $289$  & $114$   & $-5.01$ &  $ 0.64$ & $ 9.58$ & $10.04$ & $10.11$ & $ 8.87$ & $ 9.96(6)$ \cite{PRC70}\\
 \hline\hline
\end{tabular}
\end{table}

We also note that the traditional spherical magic nucleus $^{298}$Fl ($Z=114$ and $N=184$) deviates evidently from the SSL (see Fig. 1). Although the shell gap $\Delta_{\rm scale}=96.2$ MeV in $^{298}$Fl is larger than that in $^{270}$Hs and $^{298}$120, the absolute value of its shell correction ($5.16$ MeV) is obviously smaller than those of $^{270}$Hs and $^{298}$120. The inconsistency between shell gaps and shell corrections in SHN seems to imply that the deformation effect can not be neglected in the determination of magic numbers in super-heavy region. To investigate the next magic numbers beyond $Z=82$, both the shell corrections and shell gaps, and as well as the deformations should be considered, simultaneously.

\begin{center}
\textbf{IV. SUMMARY}
\end{center}

In summary, we investigated the shell corrections, shell gaps and deformations
of nuclei systematically with the latest Weizs\"acker-Skyrme (WS4) mass model. We find that the correlation between active-neutrons and
active-protons near Fermi surface might cause many nuclei ($N>Z$) with (sub)-shell closures locating along the shell stability line $N=1.37Z+13.5$.
Along this line, the double magicity of nuclei $^{46}$Si with new magic number $N=32$ and of $^{78}$Ni are
predicted, according to the corresponding shell gaps, shell corrections and nuclear deformations. For super-heavy region, the correlation influences both the shell gaps and the $\alpha$-decay energies of SHN. For the considered SHN with $116 \le Z \le 118$, the corresponding $Q_\alpha$ has the smallest value at $N=178$ rather than 184. More neutron-rich SHN such as $^{296}$118 and $^{297}$118 could be crucial and urgently required to check the reliability of the model predictions. The calculated deformation energies suggest that the slightly oblate shapes for $^{296}$118 and $^{298}$120 at their ground state provide a more stable configuration than spherical shape in the fission path.

\begin{center}
\textbf{ACKNOWLEDGEMENTS}
\end{center}

This work was supported by National Natural Science Foundation of China (Nos. 11275052, 11365005, 11422548, 11335002, 11265004) and National Key Basic Research Program of China (Grant No. 2013CB834400). N. W. thanks Zhu-Xia Li, Shan-Gui Zhou, Nikolai Antonenko and Gurgen Adamian for helpful discussions and acknowledges the support of the Open Project Program of State Key Laboratory of Theoretical Physics, Institute of Theoretical Physics, Chinese Academy of Sciences, China (No. Y4KF041CJ1). The nuclear mass tables with the WS formulas are available from http://www.imqmd.com/mass/.


\begin{thebibliography}{99}

\bibitem{Audi12} G. Audi, M. Wang, A. H. Wapstra, F. G. Kondev, M. Mac-Cormick, X. Xu, and B. Pfeiffer, Chin. Phys. C \textbf{36}, 1287 (2012).

\bibitem{Sob} A. Sobiczewski,  K. Pomorski, Prog. Part. Nucl. Phys. \textbf{58},
292 (2007).
\bibitem{Ogan} Yu. Ts. Oganessian and V. K. Utyonkov, Rep. Prog. Phys. \textbf{78}, 036301 (2015).

\bibitem{Xu13} C. Xu, X. Zhang, Z. Z. Ren, Nucl. Phys. A \textbf{898}, 24 (2013).

\bibitem{Meng} J. Meng, Y. Chen, et al., Phys. Scr. T \textbf{154}, 014010 (2013).

\bibitem{Mend14} J. Jes\'us Mendoza-Temis, G. Mart\'inez-Pinedo, et al., arXiv:1409.6135

\bibitem{Mum15} M. R. Mumpower, R. Surman, G. C. McLaughlin, A. Aprahamian, arXiv:1508.07352

\bibitem{Liu10} M. Liu, N. Wang, Z. X. Li, and F. S. Zhang, Phys. Rev. C \textbf{82}, 064306 (2010).
\bibitem{Wang15a} N. Wang,  M. Liu,  H. Jiang,  J. L. Tian, Y. M. Zhao, Phys. Rev. C \textbf{91}, 044308 (2015).
\bibitem{Zhao10} P. W. Zhao, Z. P. Li, J. M. Yao, and J. Meng, Phys. Rev. C \textbf{82},  054319 (2010).

\bibitem{Sob66} A. Sobiczewski, F.A. Gareev, B.N. Kalinkin, Phys. Lett. 22, 500 (1966).


\bibitem{Hof00} S. Hofmann and G. M\"unzenberg, Rev. Mod. Phys. \textbf{72}, 733 (2000).

\bibitem{Zeld98} N. Zeldes, Phys. Lett. B 429, 20 (1998).

\bibitem{Long14} J.J. Li, W. H. Long, J. Margueron, N. V. Giai, Phys. Lett. B \textbf{732}, 169 (2014).

\bibitem{Adam09} G.G. Adamian , N.V. Antonenko, and W. Scheid, Eur. Phys. J. A \textbf{41}, 235 (2009).


\bibitem{Lenzi} S. M. Lenzi  and M. A. Bentley, Lecture Notes in
Physics, \textbf{764}, 57 (2009).

\bibitem{Shlomo} S. Shlomo, Rep. Prog. Phys. \textbf{41}, 957 (1978)


\bibitem{Wang} N. Wang, M. Liu and X. Z. Wu, Phys. Rev. C \textbf{81}, 044322 (2010).

\bibitem{Wang10} N. Wang, Z. Y. Liang, M. Liu and X. Z. Wu, Phys. Rev. C \textbf{82}, 044304 (2010).

\bibitem{Liu11} M. Liu, N. Wang, Y. G. Deng, and X. Z. Wu, Phys. Rev. C \textbf{84}, 014333 (2011).

\bibitem{Wang14} N. Wang, M. Liu, X. Z. Wu, and J. Meng, Phys. Lett. B \textbf{734}, 215 (2014).

\bibitem{Wang11} N. Wang and M. Liu, Phys. Rev. C \textbf{84}, 051303(R)
(2011); http://www.imqmd.com/mass/

\bibitem{Niu13} Z. M. Niu, Z. L. Zhu, Y. F. Niu, et al., Phys. Rev. C \textbf{88}, 024325 (2013).

\bibitem{Sob13} A. Sobiczewski, Yu. A. Litvinov, Phys. Scr. \textbf{T154}, 014001 (2013); Phys. Rev. C \textbf{89}, 024311 (2014).
\bibitem{Lit14} Yu. A. Litvinov, M. Palczewski, E. A. Cherepanov, A. Sobiczewski, Acta Phys. Polo. B \textbf{45}, 1979 (2014).
\bibitem{Ogan15} Yu. Ts. Oganessian, V. K. Utyonkov, Nucl. Phys. A (2015), in press.

\bibitem{Cwoik} S. Cwoik, J. Dudek, W. Nazarewicz, J. Skalski, and T. Werner, Comp. Phys. Comm. \textbf{46}, 379 (1987).

\bibitem{Wien13} F. Wienholtz, D. Beck, K. Blaum, et al., Nature \textbf{498}, 346 (2013).

\bibitem{Wang15} N. Wang and M. Liu, Chin. Sci. Bull. \textbf{60}, 1145 (2015).

\bibitem{Orlandi} R. Orlandi, D. M\"ucher, R. Raabe, et al., Phys. Lett. B \textbf{740}, 298 (2015).


\bibitem{Myer66} W. D. Myers,  W. J. Swiatecki, Nucl. Phys. \textbf{81}, 1 (1966).
\bibitem{Jan05} R. Janssens, Nature \textbf{435}, 897 (2005).
\bibitem{Ang15} I. Angeli and K. P. Marinova, J. Phys. G: Nucl. Part. Phys. \textbf{42}, 055108 (2015).

\bibitem{Rutz} K. Rutz, M. Bender, T. B\"urvenich,  T. Schilling, P.-G. Reinhard, J. A. Maruhn, and W. Greiner, Phys. Rev. C \textbf{56}, 238 (1997).

\bibitem{Wang14b} Q. Mo, M. Liu, N. Wang, Phys. Rev. C \textbf{90}, 024320 (2014).


\bibitem{Ogan06} Yu. Ts. Oganessian, V. K. Utyonkov, Yu. V. Lobanov, et al., Phys. Rev. C \textbf{74}, 044602 (2006).

\bibitem{Mun03} I. Muntian, Z. Patyk, and A. Sobiczewski, Phys. Atom. Nucl. \textbf{66}, 1015 (2003).
\bibitem{Mun03a} I. Muntian, S. Hofmann, Z. Patyk, and A. Sobiczewski, Acta Phys. Pol. B \textbf{34}, 2073 (2003).


\bibitem{Kuz12} A. N. Kuzmina,  G. G. Adamian,  N. V. Antonenko,  and W. Scheid, Phys. Rev. C \textbf{85}, 014319 (2012).

\bibitem{Moll95} P. M\"oller, J. R. Nix, et al., At. Data and
Nucl. Data Tables \textbf{59} (1995) 185.


\bibitem{PRC70} Yu. Ts. Oganessian, V. K. Utyonkov, Yu. V. Lobanov, et al, Phys. Rev. C \textbf{70}, 064609 (2004).

\bibitem{PRC92} V. K. Utyonkov, N. T. Brewer, Yu. Ts. Oganessian, et al, Phys. Rev. C \textbf{92}, 034609 (2015).


\end{thebibliography}
\end{document}